\documentclass[conference]{IEEEtran}
\usepackage{amsmath}
\usepackage{algorithm}
\usepackage[noend]{algpseudocode}
\usepackage{multirow}
\usepackage{graphicx}
\usepackage{textcomp}
\usepackage{xcolor}
\usepackage{caption}
\usepackage{subcaption}
\usepackage[hidelinks]{hyperref}
\usepackage{booktabs,array}

\begin{document}

\title{FormalFixer: Intent-Aligned Automated Hardware Repair via Joint Refinement of Specifications, Assertions, and RTL Implementations}

\author{
\IEEEauthorblockN{Changyuan Yu, Hongce Zhang}
\IEEEauthorblockA{The Hong Kong University of Science and Technology (Guangzhou)\\
\texttt{cyu418@connect.hkust-gz.edu.cn}\\
\texttt{hongcezh@hkust-gz.edu.cn}}
}

\maketitle

\begin{abstract}
Functional errors in hardware designs can compromise correctness and security, and must be fixed before tape-out. Formal property verification (FPV) provides strong guarantees by proving or refuting assertions, but a failed assertion is not always evidence of an RTL defect: the failure may also come from a buggy assertion or from a specification that is semantically correct but too ambiguous or incomplete for direct formalization. This creates a practical repair problem that existing RTL-centric methods do not directly address: the repair loop must align specification intent, verification intent, and RTL behavior before deciding which artifact to change.

\textsc{FormalFixer} addresses this problem with a multi-stage, intent-aligned repair workflow for Verilog designs with assertions. It preserves the original specification as authoritative ground truth and appends confidence-filtered refinements to clarify implicit constraints. \textsc{FormalFixer} uses both intent mismatch and verifier-generated counterexample cone graphs to guide iterative RTL patching. This workflow explicitly checks specification--assertion--RTL correspondence. Results support intent alignment for FPV-based hardware repair.
\end{abstract}


\IEEEpeerreviewmaketitle

\section{Introduction}
\label{intro}
Formal property verification (FPV) exposes register-transfer level (RTL) design defects by proving or refuting assertions, but a failed assertion does not by itself identify which artifact is wrong. The root cause may be an RTL implementation bug, an incorrect assertion, or a high-level specification that is semantically correct but too implicit for direct formalization. This ambiguity makes FPV-centric repair different from conventional RTL-only repair: before patching code, the repair loop must decide how design intent, verification intent, and implementation behavior should be aligned.

\begin{figure}[!t]
\centering
\vspace{6pt}
\includegraphics[width=0.9\columnwidth]{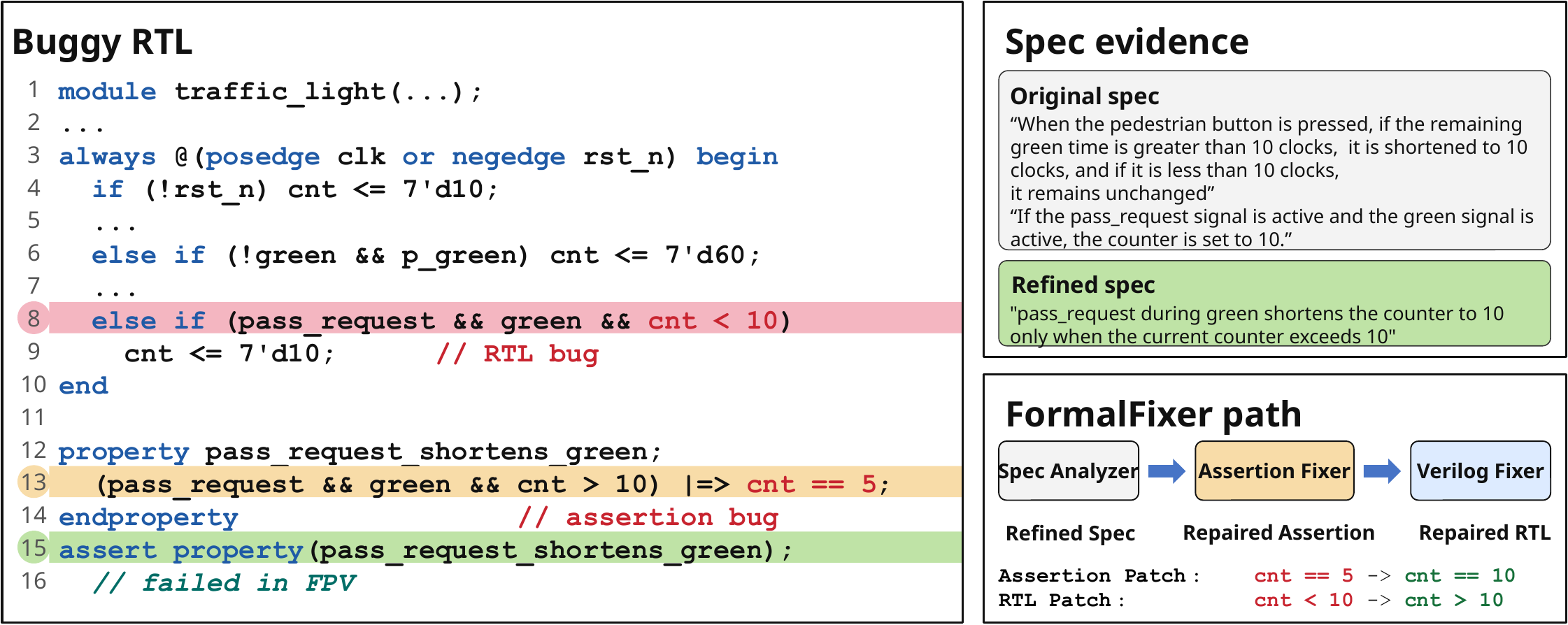}
\caption{Motivating traffic-light example linking specification refinement, assertion repair, and RTL repair.}
\label{fig:motivation-example}
\end{figure}

In Fig.~\ref{fig:motivation-example}, the traffic-light specification leaves the pedestrian-button signal and counter condition implicit. The Spec Analyzer clarifies that the button corresponds to \texttt{pass\_request}, which sets the counter to 10 during green only when its current value is below 10. This grounds both repairs in design intent rather than agreement between buggy artifacts.

This example highlights a cross-artifact repair problem: semantically correct natural language specifications may leave implementation details implicit, while assertions and RTL may both contain bugs. Existing automated repair methods do not directly address this setting.

Classical APR was first developed for software~\cite{apr-survey} and later adapted to RTL repair~\cite{cirfix1, cirfix2}.Recent LLM-based hardware repair methods further use compiler errors, tests, traces, or textual bug descriptions to generate candidate RTL patches~\cite{bugfix-cwe,hdldebugger,bugfix-rag,meic,uvllm}. These methods can reduce manual debugging effort, but they typically treat specifications and properties as correct and fixed. As a result, they are poorly matched to FPV workflows where the verification artifacts themselves may be part of the bug surface.

In this paper, we propose \textsc{FormalFixer}, an intent-driven automated hardware repair framework that jointly refines specifications, repairs assertions, and patches RTL code under intent mining and formal feedback. \textsc{FormalFixer} first appends high-confidence clarifying annotations to the original specification, repairs assertion artifacts against the refined intent, applies conservative pre-verification RTL fixes, and then uses verifier-generated counterexample cone graphs to guide iterative RTL patching. This staged decomposition makes each repair decision local to the artifact responsible for the observed mismatch. The contributions of this paper are summarized as follows:\looseness=-1
\begin{itemize}
    \item We formulate an FPV-centric hardware repair setting in which specification ambiguity, assertion bugs, and RTL defects co-exist. To our knowledge, this is the first automated hardware repair work to model this setting without assuming bug-free verification artifacts.
    \item We design \textsc{FormalFixer}, a multi-stage LLM-guided framework that preserves specification authority, appends confidence-filtered refinements, repairs assertions, and patches RTL with verifier feedback.
    \item We introduce structured counter-example (CEX) cone feedback for RTL patch synthesis. Failed properties are used to localize suspicious logic and emit small deterministic patches with rollback on failed verification.
    \item \textsc{FormalFixer} achieves 89.47\% pass@5 on \textsc{SVA-Eval-Human} and 84.46\% pass@5 on \textsc{SVA-Eval-Machine} with GLM-4.7. The direct-edit baseline without the Spec Analyzer reaches 63.16\% and 75.26\%, showing the value of explicit intent alignment.
\end{itemize}

The rest of the paper is organized as follows: Section~\ref{related} reviews related work; Section~\ref{method} presents the \textsc{FormalFixer} methodology; Section~\ref{exp} reports experimental results; and Section~\ref{conclusion} concludes the paper.

\section{Preliminaries and Related Work}
\label{related}

\subsection{SystemVerilog Assertions (SVA)}

In formal property verification (FPV), a property specification language---most commonly SystemVerilog Assertions (SVA)---encodes expected behaviors as machine-checkable properties. Assertions capture \emph{verification intent}, ranging from specification-level requirements to design-level invariants implied by \emph{design intent}. Well-structured assertions also provide rich, localized context (clocking, reset/disable conditions, diagnostics) and act as a formal interface between natural-language requirements and tool-generated verification evidence. This is central to FPV-driven debugging, because assertion quality directly influences whether a failure is attributed to an RTL defect, a property defect, or an intent mismatch across artifacts; when a property fails, the associated counterexample trace provides concrete evidence of how the RTL violates the intended behavior, which can guide intent alignment and repair.\looseness=-1

\subsection{Automated Hardware Repair}
Although software APR is well developed, automated hardware repair has received less attention. CirFix~\cite{cirfix1,cirfix2} is among the first to bring APR to hardware, using simulation-based fault localization with genetic-programming patch generation; while it can repair designs without extra information, it often needs hours on straightforward cases. Later works use LLMs with retrieved specification context~\cite{bugfix-rag}, explicit bug types and locations~\cite{bugfix-cwe}, or fine-tuning on synthetic buggy designs~\cite{hdldebugger}. However, prior works generally assume verification assets are correct and focus only on fixing RTL defects; they do not handle cases where assertions themselves are incorrect, and thus neglect collaborative repair across design and verification artifacts. \textsc{FormalFixer} addresses this gap by jointly refining intent-related verification artifacts and RTL code in an FPV-driven loop.\looseness=-1

\subsection{LLMs for EDA}

LLMs such as ChatGPT~\cite{gpt} have enabled EDA assistants that reason over natural language and source code. LLMs have been explored to generate verification artifacts (e.g., assertions from specifications~\cite{assertllm}), for hardware construction (RTL generation from natural language~\cite{mage,verigen}), and to produce or refine design specifications~\cite{specllm}. Beyond content generation, LLMs serve as interactive tool-in-the-loop systems that query EDA tools, interpret feedback, and iteratively act~\cite{chateda}. Overall, LLMs are becoming a versatile interface between human intent and EDA workflows, with growing potential in verification, debugging, and design automation.

\begin{figure*}[t]
\centering
\includegraphics[width=0.8\textwidth]{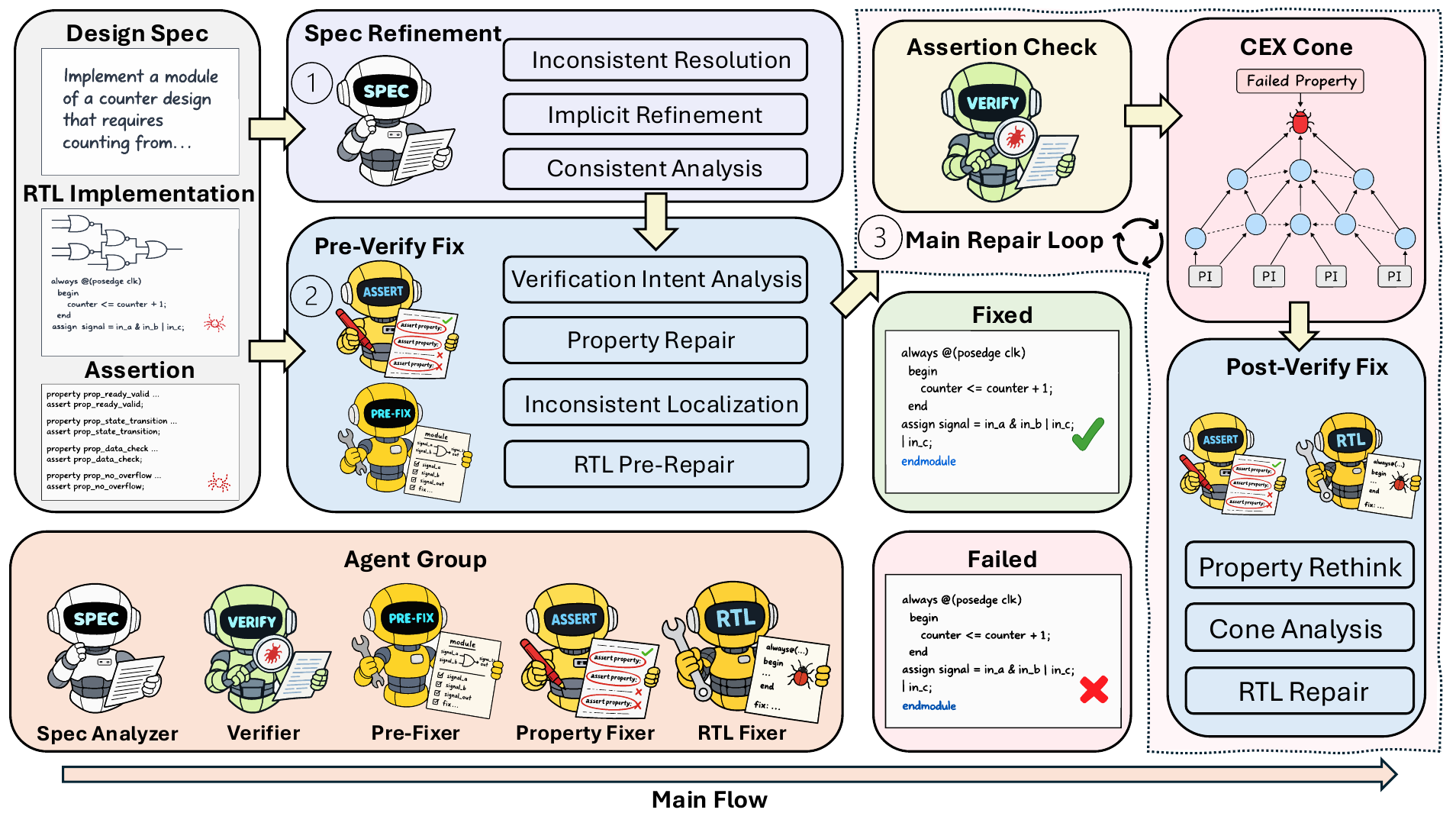}
\caption{Overall workflow of \textsc{FormalFixer}. The pipeline first aligns specification and assertion intent, then uses formal verification feedback and CEX cone analysis for iterative RTL repair.}
\label{fig:flowchart}
\end{figure*}
\section{Methodology}
\label{method}
\subsection{FormalFixer Overview}

\textsc{FormalFixer} follows an \emph{intent-aligned} workflow: it first stabilizes the relationship among specification, assertions, and RTL, and then uses formal feedback to guide localized implementation fixes. The input is a natural-language specification, Verilog RTL with embedded or associated assertions. The output is a repaired RTL/assertion pair that is rechecked by FPV.\looseness=-1

\noindent\textbf{Design motivation:} Automated RTL repair under formal verification is fundamentally challenging because a single failing property can originate from multiple coupled sources: ambiguous or underspecified natural-language requirements, buggy or misaligned assertions, and genuine implementation defects. Directly applying an LLM to patch RTL in one shot often entangles these sources, resulting in unstable edits, overfitting to current properties, and low reproducibility across runs.\looseness=-1

Our design follows three principles to address this challenge:\looseness=-1

\noindent\textbf{Preserve design intent before patching:} The original specification is treated as the authoritative high-level intent. The framework may use RTL and assertions as evidence to append clarifying refinements that specify corner-case conditions and resolve implicit or ambiguous aspects of the specification.

\noindent\textbf{Remove low-cost inconsistencies before formal-loop repair:} Assertion and simple RTL artifacts are fixed before expensive CEX-guided iterations, so the verifier is less likely to return failures caused by avoidable property defects.

\noindent\textbf{Use formal feedback for localized repair:} After FPV, we use the CEX cone graphs to localize suspicious logic and support small structured patches that can be applied and rolled back deterministically.

Fig.~\ref{fig:flowchart} illustrates the workflow. \textsc{FormalFixer} first runs a pre-verification stage consisting of the Spec Analyzer, Assertion Fixer, and Verilog Pre-Fixer. This stage aligns design and verification intent without relying on a failing CEX. The loop stage then runs the Verifier, Assertion Fixer and Verilog Fixer: failed properties are converted into cone-of-influence evidence, RTL patches are generated, and the patched design is rechecked until all properties are proven or the repair budget is exhausted.

At the core of this workflow is an explicit \emph{correspondence check} among three artifacts: specification, assertions, and RTL implementation. We treat the specification as the primary carrier of design intent, assertions as the formalization of verification intent, and RTL as the executable design realization. Repair decisions are therefore driven by identifying and reducing inconsistencies across these three views, rather than directly patching RTL from property failures alone. Concretely, we decompose the toolchain into five modules: \textbf{Spec Analyzer}, \textbf{Assertion Fixer}, \textbf{Verilog Pre-Fixer}, \textbf{Verifier}, and \textbf{Verilog Fixer}, detailed below.

\vskip -6pt
\subsection{Spec Analyzer}
\label{subsec:spec-analyzer}
\vskip -2pt

In practice, specifications may be implicit, underspecified, or partially inconsistent with the implementation, while the RTL code often relies on intermediate signals and case distinctions that are not explicitly documented. This mismatch blurs \emph{design intent} and makes it difficult to determine whether a failing property originates from an RTL defect, an assertion artifact, or an ambiguous specification.

\textbf{Ground-truth preservation.} To address this safely, the Spec Analyzer treats the \emph{original specification as the only authoritative source} throughout the pipeline. It never overwrites the original specification text; instead, it provides non-authoritative hints that clarify ambiguous or implicit expressions while preserving the original semantics. RTL and assertions are used only to expose tensions: when either artifact contradicts the original specification, the analyzer records that evidence as a conflict signal and flags the corresponding RTL/assertion behavior as suspect rather than changing the specification's semantics. Downstream fixers are instructed to ignore any refinement that conflicts with the original specification, so defects embedded in RTL or assertions are not promoted into new design intent.

\textbf{Refinement procedure.} The spec analyzer performs spec refinement by mining latent constraints and operational semantics from both RTL and assertions. Concretely, it (i) introduces and consistently names intermediate signals, (ii) formalizes and partitions operational conditions (e.g., reset versus normal operation, opcode-dependent branches, and state-dependent behaviors), and (iii) makes implicit constraints explicit (e.g., bit-width constraints, valid input ranges, and mutual-exclusion conditions). The spec analyzer highlights potential errors/ambiguities and produces a clarified, internally consistent specification that better captures the inferred design intent. \looseness-1

The Spec Analyzer outputs a refined specification that is sufficiently complete and semantically unambiguous to enable intent alignment in the subsequent repair stages.

\vskip -12pt
\subsection{Pre-verification Fixer}
\label{subsec:pre-verification-fixer}
\vskip -2pt

Before invoking formal property verification, \textsc{FormalFixer} performs a pre-verification fix step that proactively improves both the RTL code and the property set. This step consists of two complementary fixers: the Assertion Fixer repairs properties and the Verilog Pre-Fixer repairs simple RTL/spec mismatches. Both fixers consume the original specification plus accepted refinements so that property repair and RTL repair do not overwrite each other's responsibility.

\noindent\textbf{Assertion Fixer:}
The Assertion Fixer aligns verification intent with the refined design intent. It uses the refined specification and current RTL context to interpret what each property intends to check, then emits exact-site property patches in a structured JSON format with the fixed assertion, an explanation, and confidence. Typical assertion issues include incorrect temporal relationships, wrong clocking events, missing disable conditions during reset, inconsistent signal references, wrong constants, or mismatched assumptions/guarantees shown in TABLE~\ref{tab:assertion-bugs}. By repairing these issues before the main loop, later FPV failures are more likely to indicate genuine RTL defects rather than property artifacts.

\noindent\textbf{Verilog Pre-Fixer:}
Guided by the refined specification, the Verilog Pre-Fixer performs an early round of RTL code correction, targeting straightforward mistakes identifiable by directly comparing RTL against clarified requirements. Examples include missing assignments in a branch, wrong conditional guards, incorrect reset behavior, or obvious spec-mismatching expressions. The module emits mini-patches with corrected RTL snippets, locations, rationales, and confidence. To remain conservative, only the highest-confidence accepted patch is applied in this stage. This produces a cleaner starting point for formal runs and reduces the search space for later CEX-guided repair.

\vspace{-4pt}

\subsection{Verifier}
\label{subsec:verifier}

The Verifier runs FPV on the current RTL--property pair and converts failures into structured repair evidence. It invokes a model checker and collects the status of each property as proven, failed, or unknown. When an assertion fails, the verifier retrieves the corresponding counterexample and constructs a cone-of-influence (COI) graph: starting from signals referenced in the failed property, it recursively traces predecessors through the RTL until reaching primary inputs or the extraction budget. The graph is serialized as JSON with \texttt{nodes}, \texttt{edges}, and \texttt{cycles}; \texttt{nodes} identify relevant signals or expressions, \texttt{edges} encode dependencies, and \texttt{cycles} record cycle-indexed signal/value evidence from the CEX. If all properties are proven, the workflow terminates; otherwise, property-related failures can re-trigger assertion repair, while implementation-related failures guide the Verilog Fixer.

When the property fails, which means the RTL implementation or assertion still mis-aligned with the design intent and verification intent, a CEX-guided repair loop will be used to further repair the potential bug. At each round, the assertion fixer is first invoked to check if the verification intent matches the design intent. If the buggy assertion cause the verification fails, we first try to fix the assertion, and then invokes Verilog Fixer.

\vspace{-4pt}
\subsection{Verilog Fixer}
\label{subsec:verilog-fixer}

The Verilog Fixer consumes the refined specification, current failed/unknown properties, the CEX cone graph, and previous-round patch context as input. The CEX can clearly show a trace that cause the property failed, by analyzing the CEX trace and align it with the design intent, it can infer whether the bug exists in assertions, RTL or both. The previous patches will also be given to the LLM to avoid rollback. As the previously invoked Assertion Fixer already diagonise the assertion, Verilog Fixer then trys to generate new patches to repair the potential buggy RTL. If the failure is caused by the assertion only, none RTL patch will be applied. After each round, the patched RTL is reverified. If verification fails because the current-round patches broke elaboration or proof setup, \textsc{FormalFixer} rolls back to the last verified checkpoint for that round and continues from that stable state. If verification succeeds but properties remain failed or unknown, the new CEX cone and previous patch rationale are fed into the next round. The loop repeats until all properties are proven or the three-round repair budget is reached.

\section{Experiment}
\label{exp}

\begin{table}[!t]
\centering
\setlength{\tabcolsep}{4pt}
\renewcommand{\arraystretch}{1.03}
\footnotesize
\caption{Examples of common assertion bugs.}
\
\label{tab:assertion-bugs}
\resizebox{0.9\linewidth}{!}{
\begin{tabular}{@{}>{\centering\arraybackslash}m{0.24\linewidth}>{\centering\arraybackslash}m{0.70\linewidth}@{}}
\toprule
\multicolumn{1}{c}{\textbf{Error}} & \multicolumn{1}{c}{\textbf{Symptoms}} \\
\midrule

\multirow{2}{*}{\centering Polarity Error}
& \textcolor{red}{\texttt{RST |=> valid\_out == 0;}} \\
& \textcolor{green!50!black}{\texttt{RST |=> valid\_out == 1;}} \\
\midrule

\multirow{2}{*}{\centering Bitwidth Misuse}
& \textcolor{red}{\texttt{cout == (a+b+cin > 256);}} \\
& \textcolor{green!50!black}{\texttt{cout == (a+b+cin > 255);}} \\
\midrule

\multirow{2}{*}{\centering Mapping Error}
& \textcolor{red}{\texttt{(a+b+Cin) == \{y, Co\};}} \\
& \textcolor{green!50!black}{\texttt{(a+b+Cin) == \{Co, y\};}} \\
\midrule

\multirow{2}{*}{\centering Operator Error}
& \textcolor{red}{\texttt{res == a\_signed - b\_signed;}} \\
& \textcolor{green!50!black}{\texttt{res == a\_signed + b\_signed;}} \\
\midrule

\multirow{2}{*}{\centering Edge Error}
& \textcolor{red}{\texttt{\$rose(CLK) |=> \$rose(CLK);}} \\
& \textcolor{green!50!black}{\texttt{\$rose(CLK) |=> \$fell(CLK);}} \\
\midrule

\multirow{2}{*}{\centering Temporal Error}
& \textcolor{red}{\texttt{res == \$past(a,3)+\$past(b,3);}} \\
& \textcolor{green!50!black}{\texttt{res == \$past(a,4)+\$past(b,4);}} \\
\midrule

\multirow{2}{*}{\centering Signal Misuse}
& \resizebox{\linewidth}{!}{\textcolor{red}{\texttt{(state==IDLE)\&\&en |=> (state==SHIFT);}}} \\
& \textcolor{green!50!black}{\texttt{(state==IDLE)\&\&en |=> (state==SUB);}} \\
\bottomrule
\end{tabular}}
\end{table}

\begin{table*}[ht]
\centering
\setlength{\tabcolsep}{4pt}
\renewcommand{\arraystretch}{1.08}
\footnotesize
\caption{Overall repair effectiveness on \textsc{SVA-Eval-Human} and \textsc{SVA-Eval-Machine}.}
\label{tab:rq1-prelim-results}
\begin{tabular}{@{}l l l c c c c c c c c@{}}
\toprule
\textbf{SA} & \textbf{Model} & \textbf{Fixer} & \multicolumn{4}{c}{\textbf{\textsc{SVA-Eval-Human} (\%)}} & \multicolumn{4}{c}{\textbf{\textsc{SVA-Eval-Machine} (\%)}} \\
\cmidrule(lr){4-7}\cmidrule(lr){8-11}
& & & \textbf{Pass@1} & \textbf{RTL Fixed@1} & \textbf{Pass@5} & \textbf{RTL Fixed@5} & \textbf{Pass@1} & \textbf{RTL Fixed@1} & \textbf{Pass@5} & \textbf{RTL Fixed@5} \\
\midrule
\multirow{4}{*}{Yes} & Qwen3.5-Flash & Separate & 68.42 & 78.95 & 84.21 & 92.11 & 72.57 & 73.37 & 82.86 & 84.01 \\
& GLM-4.7 & Separate & 78.95 & 92.11 & 89.47 & 94.74 & 76.23 & 77.14 & 84.46 & 85.14 \\
& DeepSeek V3.2 & Separate & 71.05 & 84.21 & 86.84 & 94.74 & 71.02 & 72.39 & 82.01 & 82.57 \\
& GLM-4.7 & Shared & 42.11 & 73.68 & 68.42 & 84.21 & 59.03 & 64.80 & 76.97 & 74.17 \\
\midrule
\multirow{2}{*}{No} & GLM-4.7 & Separate & 63.16 & 84.21 & 68.42 & 71.05 & 63.31 & 73.94 & 70.06 & 73.26 \\
& GLM-4.7 & Shared & 42.11 & 76.32 & 63.16 & 89.47 & 60.08 & 68.79 & 75.26 & 77.82 \\
\bottomrule
\end{tabular}
\end{table*}

\subsection{Benchmark Preparation}

We build our benchmark on existing open-source RTL repair datasets~\cite{meic,uvllm}, which are constructed from real-world RTL designs~\cite{rtllm} with representative functional bugs injected. We extend these cases with assertion bugs to evaluate joint repair. The final benchmark contains 38 \textsc{SVA-Eval-Human} cases and 875 \textsc{SVA-Eval-Machine} cases, each containing RTL code, assertions, and a natural-language specification.

The assertion bugs are generated with controlled templates rather than free-form fuzzing. Specifically, we define a few-shot template for each category in our assertion-bug taxonomy (Table~\ref{tab:assertion-bugs}), grounded in common assertion-error types and patterns. To avoid knowledge leakage, GPT-4o is used only to instantiate these templates on existing assertions; it is not used as one of the repair models. This construction makes the generated assertion bugs \emph{taxonomically grounded}, \emph{controlled in diversity}, and \emph{reproducible}. We retain the original imperfect specifications to preserve realism and avoid additional human bias in the repair target.

\subsection{Experimental Setup}

To evaluate robustness across foundation models, we run \textsc{FormalFixer} with DeepSeek V3.2~\cite{deepseek-v3-techreport}, GLM-4.7~\cite{glm-47-release}, and Qwen3.5-Flash~\cite{qwen3-techreport}. The three models are in a similar API-pricing range, and the Flash-tier model tests whether a lightweight model can still handle intent alignment and verifier feedback. All models are accessed via OpenRouter~\cite{openrouter-api}, with reasoning disabled and default generation parameters.

We ablate two factors: \textbf{Spec Analyzer} indicates whether the module is enabled (Yes/No), and \textbf{Fixer} indicates whether assertions and RTL are edited in separate LLM calls (Separate) or a single call (Shared). All configurations retain CEX-guided iteration. The default configuration uses the Spec Analyzer and separate fixers; GLM-4.7 is evaluated in all four combinations. \emph{Plain-GLM-4.7} in Fig.~\ref{fig:rq4-complexity} denotes shared editing without the Spec Analyzer.

We use Cadence JasperGold\textsuperscript{\textregistered} for FPV and leverage its \texttt{visualize -why} and \texttt{visualize -value} commands to build CEX cone graphs. CEX extraction is capped at 20 time frames and 200 nodes, and the repair-round budget is 3.

Prior work typically assumes bug-free assertions, whereas our setting explicitly includes assertion bugs. Therefore, direct comparison with UVLLM~\cite{uvllm}, MEIC~\cite{meic}, AssertSolver~\cite{assertsolver}, and FVDebug~\cite{fvdebug} is not fully meaningful: UVLLM and MEIC repair RTL under fixed verification assets, while AssertSolver and FVDebug analyze assertion failures under the assumption that the failed property represents the intended behavior. In our setting, the assertion itself may be wrong; treating it as an oracle can push a repair method to modify RTL toward a faulty property or to miss the need for assertion repair. We nevertheless discuss these methods as reference points for RTL-centric repair performance.

To support reproducibility, we provide an anonymous artifact release at \url{https://anonymous.4open.science/r/formalfixer-open-70B3/}, which includes module prompts, benchmark artifacts, and the verifier-backend interface. We do not release Cadence JasperGold-related integration code because of licensing and IP restrictions; users may instead plug in their own JasperGold setup or substitute another verifier that implements the same interface.

\subsection{Results and Ablation Study}

We organize the analysis around four research questions.

\textbf{RQ1: How effective is \textsc{FormalFixer} on \textsc{SVA-Eval-Human} and \textsc{SVA-Eval-Machine}?}

We report two metrics. \emph{Pass@k} counts a case as solved when both RTL and assertions are correct within $k$ attempts. \emph{RTL Fixed@k} counts a case as solved when the RTL functionality is corrected, regardless of whether the generated assertion repair is also correct. We also inspect whether a patch changes semantics outside the buggy region, which helps identify false negatives. Results are shown in Table~\ref{tab:rq1-prelim-results}.

Across models, RTL Fixed@1 is higher than Pass@1, suggesting that RTL--spec alignment is easier for LLMs than assertion--spec alignment. Writing correct SVA requires specialized semantics that may be less represented in general-purpose training data, so the RTL can be fixed while the paired assertion remains imperfect. Since the evaluated models were released after the public source benchmarks, training-data exposure may also contribute to this gap.

With the default configuration, GLM-4.7 achieves the highest Pass@1 and Pass@5 on both datasets, reaching 89.47\% and 84.46\% Pass@5 on Human and Machine, respectively. Its corresponding RTL Fixed@5 rates are 94.74\% and 85.14\%. Qwen3.5-Flash remains competitive at 84.21\% and 82.86\% Pass@5, while DeepSeek V3.2 also benefits from multiple attempts.

The default GLM-4.7 configuration outperforms all three ablations in Pass@5. Shared editing without the Spec Analyzer achieves 63.16\% on Human and 75.26\% on Machine. Its Human RTL Fixed@5 rate of 89.47\% substantially exceeds Pass@5, indicating that RTL correction often leaves assertions incorrect. RQ2 examines the contributions of the Spec Analyzer and fixer organization separately.

The Human--Machine comparison further clarifies where the difficulty arises. The performance drop is more visible in later-round RTL-fix behavior than in first-attempt pass behavior, suggesting that the challenge is not merely locating the initial buggy region. Harder cases require the repair trajectory to preserve a consistent relationship among the specification, repaired assertions, and RTL patches across multiple iterations. Once this relationship drifts, later verifier feedback may guide local fixes without ensuring global correctness.

\begin{table*}[!t]
\centering
\setlength{\tabcolsep}{4pt}
\renewcommand{\arraystretch}{1.08}
\footnotesize
\caption{Stage-wise repair and reliability across ablation configurations.}
\label{tab:rq23-merged-results}
\begin{tabular}{@{}l l l c c c c c c@{}}
\toprule
\textbf{SA} & \textbf{Model} & \textbf{Fixer} & \multicolumn{3}{c}{\textbf{\textsc{SVA-Eval-Human} (\%)}} & \multicolumn{3}{c}{\textbf{\textsc{SVA-Eval-Machine} (\%)}} \\
\cmidrule(lr){4-6}\cmidrule(lr){7-9}
& & & \textbf{Pre-Fix@5} & \textbf{Staged Fix@5} & \textbf{False-Neg@5} & \textbf{Pre-Fix@5} & \textbf{Staged Fix@5} & \textbf{False-Neg@5} \\
\midrule
\multirow{4}{*}{Yes} & Qwen3.5-Flash & Separate & 50.00 & 34.21 & 2.63 & 51.20 & 34.06 & 6.47 \\
& GLM-4.7 & Separate & 57.89 & 31.58 & 2.63 & 43.09 & 41.37 & 5.43 \\
& DeepSeek V3.2 & Separate & 55.26 & 31.58 & 0.00 & 46.97 & 34.74 & 5.49 \\
& GLM-4.7 & Shared & 28.95 & 39.47 & 0.00 & 49.14 & 27.83 & 10.51 \\
\midrule
\multirow{2}{*}{No} & GLM-4.7 & Separate & 31.58 & 36.84 & 2.63 & 41.71 & 28.35 & 12.34 \\
& GLM-4.7 & Shared & 34.21 & 31.58 & 0.00 & 44.80 & 24.57 & 9.94 \\
\bottomrule
\end{tabular}
\end{table*}

\begin{figure*}[t]
\centering
\begin{subfigure}{0.48\textwidth}
\centering
\includegraphics[width=\linewidth]{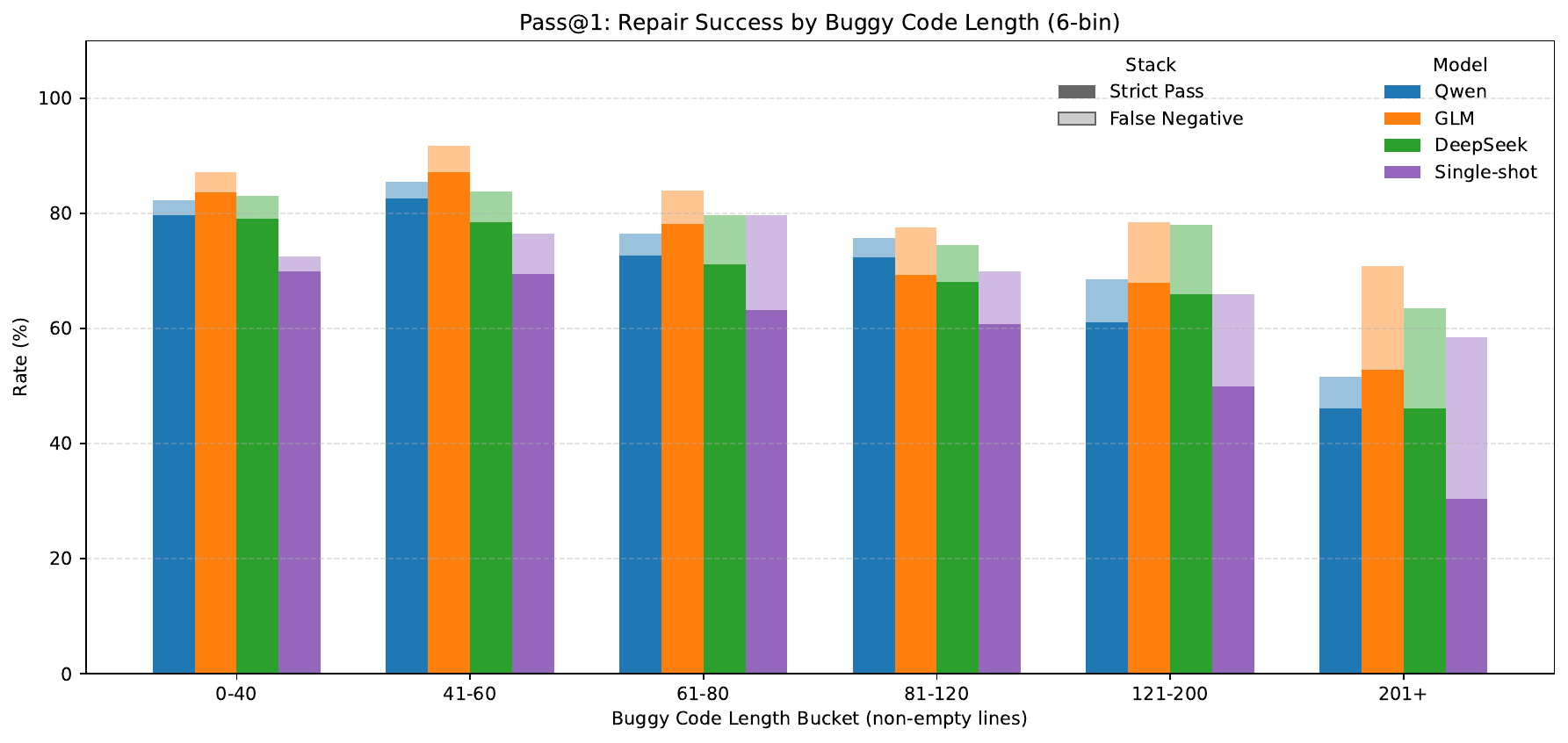}
\caption{Pass@1 across case complexity}
\label{fig:rq4-bar-pass1}
\end{subfigure}
\hfill
\begin{subfigure}{0.48\textwidth}
\centering
\includegraphics[width=\linewidth]{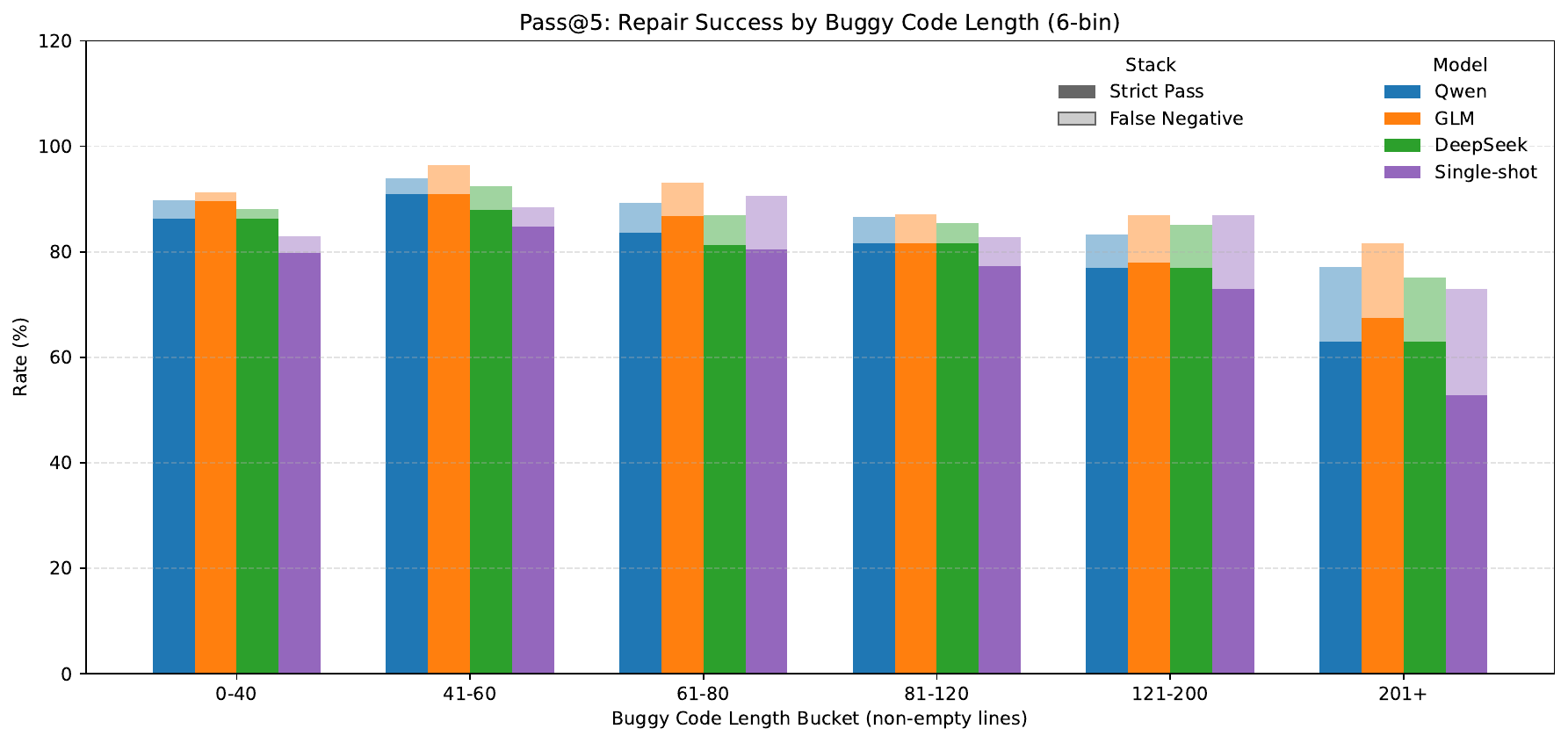}
\caption{Pass@5 across case complexity}
\label{fig:rq4-bar-pass5}
\end{subfigure}
\caption{Repair success by RTL code length. Longer RTL complicates intent recovery and localization, favoring staged alignment.}
\label{fig:rq4-complexity}
\end{figure*}

\textbf{RQ2: How do specification analysis and fixer organization affect repair?}

The GLM-4.7 ablations in Table~\ref{tab:rq1-prelim-results} vary the Spec Analyzer and fixer organization independently. With separate fixers, the Spec Analyzer improves Pass@5 by 21.05 percentage points on Human and 14.40 on Machine. With the Spec Analyzer enabled, separate editing outperforms shared editing by 21.05 and 17.49 points, respectively. Neither factor is uniformly beneficial alone: on Machine, adding the Spec Analyzer to shared editing reduces Pass@5 from 75.26\% to 66.97\%, while separate editing without the Spec Analyzer achieves only 70.06\%. These results suggest that specification analysis and separate repair are most effective in combination.

Table~\ref{tab:rq23-merged-results} reports \emph{Pre-Fix@5}, the success rate before verification; \emph{Staged Fix@5}, the additional success rate from CEX-guided repair; and \emph{False-Neg@5}, the rate of verifier-passing cases that fail stronger validation. For the default GLM-4.7 configuration, pre-verification repair resolves 57.89\% of Human and 43.09\% of Machine cases, with CEX-guided repair contributing another 31.58 and 41.37 percentage points. On Machine, shared editing with the Spec Analyzer has higher Pre-Fix@5 (49.14\%) but lower CEX-guided recovery (17.83 points), indicating that early repair success does not necessarily translate into stronger final performance.

\textbf{RQ3: Are there false-negative cases during the repair process, and how frequently do they occur?}

A case is a \emph{false negative} when it passes the available checks but is later found incorrect under stronger validation. This can happen when an assertion is weakened to accommodate buggy RTL, or when verification intent is misinterpreted so that the true bug is no longer triggered. Such cases are especially important for FPV-centric repair because a passed property can otherwise give a false sense of correctness.

False negatives concentrate in hard instances where bug localization and intent alignment are tightly coupled. In the first situation, the specification is often implicit or underspecified while the RTL logic has interacting constraints, such as temporal semantics, reset/clock behavior, or cross-signal dependencies. The repair model can then drift toward assertion-side accommodation instead of true RTL correction. In the second situation, a badly designed property can make verification intent ambiguous, allowing the model to produce a locally consistent but semantically weaker property.

On Machine, the default GLM-4.7 configuration has the lowest False-Neg@5 (5.43\%). With separate fixers, enabling the Spec Analyzer reduces False-Neg@5 from 12.34\% to 5.43\%, suggesting that intent alignment substantially improves repair reliability. This benefit is not uniform: with shared editing, the rate changes only from 9.94\% to 10.51\%, while Human results remain 2.63\% for separate fixers and 0.00\% for shared fixers regardless of the Spec Analyzer. Manual failed-case inspection indicates that hard cases are often not pure localization failures: the initial suspicious region is usually reasonable, but early partial fixes can leave assertions and RTL misaligned. Later rounds then spend their budget restoring local consistency rather than correcting global behavior, highlighting the need for stable intent alignment across rounds.

\textbf{RQ4: How does performance scale with case complexity, measured by RTL code length?}

We use RTL code length as a practical proxy for case complexity and stratify cases into length buckets. Fig.~\ref{fig:rq4-complexity} shows a clear complexity-dependent trend: performance is strongest in the small-to-medium range and declines as RTL grows. Very short fragments sometimes provide too little structural context for reliable intent recovery, whereas moderate-length designs expose enough control and dataflow structure to stabilize specification--assertion--RTL alignment. Once the design becomes larger, the search space for localization and patch validation expands, and long-range dependencies make iterative repair more fragile.

False negatives also become more frequent as RTL grows. Longer designs tend to contain more implicit behaviors and long-range dependencies, so a locally plausible patch is more likely to satisfy the current properties while deviating from the original requirement. The pass@1--pass@5 comparison further illustrates this trend: on simpler cases, repeated attempts bring limited change because the initial repair decision is already relatively stable; on harder long-code cases, repeated attempts recover part of the lost performance but cannot eliminate false negatives from persistent intent ambiguity.

Among the plotted configurations, default GLM-4.7 is most robust in the medium and long buckets, with a widening gap over Plain-GLM-4.7. This comparison reflects the combined effect of the Spec Analyzer and separate editing, rather than either factor alone. Qwen3.5-Flash remains competitive on shorter cases but is more sensitive to longer designs, where repair requires coordinating non-local RTL behavior with specification intent and verifier feedback. CPU-core-level designs remain beyond the scope of the current evaluation.

\vspace{-5pt}
\section{Conclusion}
\label{conclusion}

This paper presents \textsc{FormalFixer}, an intent-aligned framework for FPV-centric hardware repair where specification ambiguity, assertion bugs, and RTL defects may coexist. By preserving the original specification, appending confidence-filtered refinements, repairing assertions, and using CEX-cone feedback for RTL patching, \textsc{FormalFixer} separates repair decisions across artifacts. Experiments on \textsc{SVA-Eval-Human} and \textsc{SVA-Eval-Machine} show higher final pass rates and less semantic drift than direct editing without specification refinement.

\textbf{Limitations and future work.} Our evaluation focuses on block-level RTL with localized bugs and templated assertion errors, which does not fully capture large industrial designs, multi-bug interactions, or long-horizon architectural defects. Future work will extend the benchmark to larger designs and industrial assertion-error corpora.

\bibliographystyle{IEEEtran}
\bibliography{ref}
\end{document}